\documentclass[11pt]{article}

 \hoffset= - 0.75 in

\usepackage[dvips]{graphicx,color}

\usepackage{latexsym,array,delarray,amsthm,amssymb}
\usepackage[sumlimits]{amsmath}
\usepackage{color}
 \usepackage{url}

\ifx\pdftexversion\undefined
  \usepackage[dvips]{graphicx}
\else
  \usepackage[pdftex]{graphicx}
\fi
 
\newtheorem{thm}{Theorem}

\newtheorem{algorithm}[thm]{Algorithm}

\theoremstyle{definition}

\theoremstyle{remark}

\begin{document}

\title{Maximum likelihood estimation of phylogenetic tree and 
substitution rates via generalized neighbor-joining and the EM algorithm}

\author{  Asger Hobolth  \\ Bioinformatics Research Center \\  North Carolina 
   State University \\ Raleigh, NC 27606 \\
   asger@statgen.ncsu.edu \and Ruriko
  Yoshida \\ Department of Mathematics \\ Duke University \\ Durham,
  NC 27708\\ ruriko@math.duke.edu}

\maketitle

\vskip 0.2in

\section*{Abstract}
A central task in the study of molecular sequence data from present-day
species is the reconstruction of the ancestral relationships.
The most established approach to tree reconstruction is the maximum likelihood
(ML) method. In this method, evolution is described in terms of a 
discrete-state continuous-time Markov process on a phylogenetic tree.
The substitution rate matrix, that determines the Markov process, can be 
estimated using the expectation maximization (EM) algorithm.
Unfortunately, an exhaustive search for the ML phylogenetic tree
is computationally prohibitive for large data sets. In such situations,
the neighbor-joining (NJ) method is frequently used because of its 
computational speed. The NJ method reconstructs trees by clustering 
neighboring sequences recursively, based on pairwise comparisons
between the sequences. The NJ method can be generalized such that 
reconstruction is
based on comparisons of subtrees rather than pairwise distances.
In this paper, we present an algorithm for simultaneous 
substitution rate estimation and phylogenetic tree reconstruction. 
The algorithm iterates between the EM algorithm for estimating substitution 
rates and the generalized NJ method for tree reconstruction. 
Preliminary results of the approach are encouraging.
\section{Introduction}
Current efforts to reconstruct the tree of life for different organisms
demand the inference of phylogenies from thousands of DNA sequences.
The first author is at a university where the tree of life for flies 
is investigated (http://www.inhs.uiuc.edu/cee/FLYTREE/) and the second
author is at a university where the tree of life for fungi is considered
(http://ocid.nacse.org/research/aftol/). For such large data sets
the tree space is enormous and identification of the optimal tree
is a major challenge. 

The evolution of homologous DNA sequences can be described by continuous 
time Markov chains on a phylogenetic tree \cite{Galtier2005}. 
A continuous time Markov chain
is characterized by a substitution rate matrix, and the phylogenetic tree 
summarizes the relationships between the species in terms of edge lengths 
(times since divergence) and common ancestors. The DNA sequences are only 
observed in the leaves, and information on the phylogenetic tree, 
substitution events (time and type) and edge lengths is missing.
The transition matrix $P(t)$ for a continuous time Markov process can be 
written as $\exp(Qt)$, where $Q$ is a parameterized substitution rate matrix.
In order to estimate the rate parameters and edge lengths for a fixed tree
and based on the observed data, one can use the expectation maximization (EM) 
algorithm \cite{Hobolth2005}. 
The updating step of the EM algorithm can be written explicitly in 
terms of eigenvalues and eigenvectors of~$Q$.
The continuous time Markov process gives rise to a distance measure between
any sets of sequences. Pairwise distances can be used, together with the 
neighbor joining (NJ) algorithm, to reconstruct the phylogenetic tree that
relates the sequences. The generalized neighbor joining (GNJ) algorithm 
\cite{Levy2005} improves the NJ algorithm by using 
distance measures based on subtrees rather than pairwise distances.

In this paper, we describe an algorithm that simultaneously estimates the
substitution rate matrix and reconstructs the phylogenetic tree. The algorithm
iterates between the EM algorithm for rate matrix estimation and the 
GNJ algorithm for phylogenetic tree reconstruction. We are in the process of
implementing the algorithm, and preliminary results are encouraging.   
\section{The EM algorithm}
\subsection{Two sequences} 
Consider the somewhat unreasonable situation where we have 
access to the {\em complete observation} (the {\em hidden model} or the 
{\em complete data model}) of the evolution of a single site.
The state of the process at time $t$ is denoted $x(t)$, and we have 
observed the process from time $t=0$ to time $t=T$.
In this paper the size $\alpha$ of the state space is 4 corresponding to the 
four nucleotides $\Sigma = $\{{\tt A,G,C,T}\}, but our method can also be 
applied to
protein-coding sequences where the state space is of size 61 because there
are 61 sense codons.
Continuous-time Markov processes are described in e.g. \cite{Guttorp1995}.

We now derive the likelihood for a complete observation of a continuous-time
Markov process with substitution rate matrix $Q$.
Suppose the evolution of a single site has been completely observed
from time $t=0$ to time $t=T$ and is given as in Fig.~1. 
We model the evolution in terms of a continuous-time Markov process with 
substitution rate matrix $Q$. Recall that a rate matrix has non-negative 
off-diagonal entries and each row sums to zero.  Let $Q(a,b)$ be the 
entry of $Q$ in the $a$th column and the $b$th row.
The waiting time in a state $a$ is exponentially distributed with parameter 
$-Q(a,a)$ and the probability of substituting $a$ with $b$ is proportional to
$Q(a,b)$. Thus, the likelihood for the complete observation in Fig.~1 is 
given by 
\begin{eqnarray*}
  L(Q)&=&Q(1,1) e^{Q(1,1)t_1} \frac{Q(1,3)}{Q(1,1)} 
       Q(3,3) e^{Q(3,3)(t_2-t_1)} \frac{Q(3,1)}{Q(3,3)}
       Q(1,1) e^{Q(1,1)(t_3-t_2)} \frac{Q(1,2)}{Q(1,1)}
       e^{Q(2,2)(T-t_2)} \\
      &=& e^{Q(1,1)(t_1+(t_3-t_2))+Q(2,2)(T-t_2)+Q(3,3)(t_2-t_1)}
       Q(1,2) Q(1,3) Q(3,2),
\end{eqnarray*}
where indices 1,2,3,4 corresponds to {\tt A},{\tt G},{\tt C},{\tt T}.

\begin{figure}[!htb]
  \centering
  \setlength{\unitlength}{0.08mm}
  \begin{picture}(1000,700)
    \thicklines
    \put(100,100){\vector(1,0){1000}} \put(1100,50){\makebox(0,0){Time}}
    \put(100,80){\line(0,1){40}}      \put(100,50){\makebox(0,0){$0$}}
    \put(900,80){\line(0,1){40}}      \put(900,50){\makebox(0,0){$T$}}
    \put(300,80){\line(0,1){40}}      \put(300,50){\makebox(0,0){$t_1$}}
    \put(400,80){\line(0,1){40}}      \put(400,50){\makebox(0,0){$t_2$}}
    \put(700,80){\line(0,1){40}}      \put(700,50){\makebox(0,0){$t_3$}}
    \put(100,100){\line(0,1){600}} \put(0,670){\makebox(0,0){State}}
    \put(80,250){\line(1,0){40}} \put(50,250){\makebox(0,0){\tt A}}
    \put(80,350){\line(1,0){40}} \put(50,350){\makebox(0,0){\tt G}}
    \put(80,450){\line(1,0){40}} \put(50,450){\makebox(0,0){\tt C}}
    \put(80,550){\line(1,0){40}} \put(50,550){\makebox(0,0){\tt T}}
    \put(100,250){\circle*{20}}
    \put(100,250){\line(1,0){200}}
    \put(300,250){\circle{20}}
    \put(300,450){\circle*{20}}
    \put(300,450){\line(1,0){100}}
    \put(400,450){\circle{20}} 
    \put(400,250){\circle*{20}}
    \put(400,250){\line(1,0){300}} 
    \put(700,250){\circle{20}}
    \put(700,350){\circle*{20}}
    \put(700,350){\line(1,0){200}}
    \put(900,100){\line(0,1){600}}
  \end{picture}
  \caption{Complete observation of the evolution of a single site in a 
    DNA sequence.}
\end{figure}
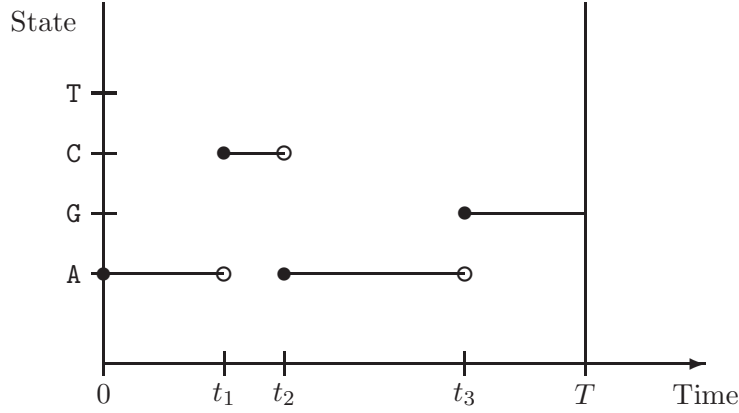
\vspace{-3mm} 
More generally, if the rate matrix is parametrized by $\theta,\; Q=Q_{\theta}$,
and with $x=\{x(t):0\leq t \leq T\}$, the maximum likelihood estimation 
problem for a complete observation is:
\begin{equation}
 {\rm maximize} \quad  L(\theta; x) \,\, = \,\,
 \left[\prod_{a \in \Sigma} \prod_{b \not = a} 
   f_{ab}(\theta)^{N(a,b)}\right]
 \left[ \prod_{a \in \Sigma} 
   f'_a(\theta)^{T(a)}\right] \quad \,\,\, 
 \hbox{subject to} \quad \theta \in \Theta.
\end{equation}
Here $f_{ab}(\theta)$ and $f'_a(\theta)$ are functions from 
${\mathbb R}^d \to {\mathbb R}_+ = \{x \in {\mathbb R}: x > 0\}$ such that 
$f_{ab}(\theta) = Q_{\theta}(a,a), \, 
f'_{a}(\theta) = \exp(Q_{\theta}(a,a)) \mbox{ for } \theta \in 
\Theta \subset {\mathbb R^d}$, $T(a)$ is the total time spent in 
state $a$, and $N(a, b)$ is the number of substitutions of $a$ with $b$. 
Thus, the log-likelihood for a complete observation becomes
\begin{eqnarray}
  \log L(\theta;x)=
  \sum_{a\in \Sigma} T(a)Q_{\theta}(a,a)+\sum_{a\in \Sigma}
  \sum_{b \neq a} N(a,b) \log Q_{\theta}(a,b).
  \label{CompleteLogL}
\end{eqnarray}
Note that the complete log-likelihood is linear in the total time spent in 
a state and the number of substitutions between states.
The complete log-likelihood function is analytically tractable. 
Indeed, as argued in \cite{Hobolth2005}, in most Markov processes of 
sequence evolution the maximum likelihood estimates can be found analytically 
from a complete observation. 

Consider for example the general time reversible (GTR) model.
Let $\pi_a,\; a\in \Sigma, \; \sum_a \pi_a=1,$ denote the stationary 
distribution of the Markov chain. This distribution can be estimated from 
the nucleotide frequencies in a single sequence.
The GTR model has substitution rate matrix (e.g., \cite{Yap2004})
\begin{eqnarray}
  Q_{\theta}= \left[ \begin{array}{cccc}
    \cdot & \theta_{12}\pi_2 & \theta_{13}\pi_3 & \theta_{14}\pi_4 \\
    \theta_{12}\pi_1 & \cdot & \theta_{23}\pi_3 & \theta_{24}\pi_4 \\
    \theta_{13}\pi_1 & \theta_{23}\pi_2 & \cdot & \theta_{34}\pi_4 \\
    \theta_{14}\pi_1 & \theta_{24}\pi_2 & \theta_{34}\pi_3 & \cdot \\
  \end{array} \right]
  \label{GTR}
\end{eqnarray}
where the diagonal elements are such that each row sums to zero and
the 6 unknown parameters are
$\theta=(\theta_{12},\theta_{13},\theta_{14},
         \theta_{23},\theta_{24},\theta_{34})$.
A simple calculation shows that the complete log-likelihood 
(\ref{CompleteLogL}) is maximized for
\begin{eqnarray*}
  \theta_{ab}^*=\frac{N(a,b)+N(b,a)}{\pi_b T(a)+\pi_a T(b)}, \;\; a < b.
\end{eqnarray*}

The problem is, however, that the DNA sequences are only
observed in the leaves, whereas information on substitution events
(time and type) and edge lengths is missing. For two sequences
we only observe the beginning state~$x(0)$ and the final state~$x(T)$ of the 
process. We have a {\em missing data problem} in the sense that we only have 
access to part of the data.

The Expectation Maximization (EM) algorithm \cite{Dempster1977}
is a broadly applicable approach for missing data problems.
The algorithm is an iterative procedure with two steps in each iteration,
the Expectation Step (E-step) and the Maximization Step (M-step).
The algorithm approaches the problem of solving the actually observed
log-likelihood indirectly by proceeding iteratively in terms of the
complete log-likelihood.
As the complete log-likelihood is unobservable, it is replaced by its
conditional expectation given the observed data $y=\{x(0),x(T)\}$.
In the E-step of the $(k+1)$'th iteration, the function
$  G(\theta;\theta^k)=
  {\rm E}_{\theta^k}[\log L(\theta;x)|y]$
is calculated, and in the M-step a new parameter value $\theta^{k+1}$ is
obtained as the value of $\theta$ that maximizes $G(\theta;\theta^k)$.
For the GTR model (\ref{GTR}) the M-step becomes
\begin{eqnarray*}
  \theta_{ab}^*=\frac{{\rm E}_{\theta^k}[N(a,b)|x(0),x(T)]+
                    {\rm E}_{\theta^k}[N(b,a)|x(0),x(T)]}
    {\pi_b{\rm E}_{\theta^k}[T(a)|x(0),x(T)]+
     \pi_a{\rm E}_{\theta^k}[T(b)|x(0),x(T)]}, \;\; a < b.
\end{eqnarray*}
The algorithm converges to a local maximum likelihood estimate of the observed
data.
 
Since the complete log-likelihood is linear in the time spent in
a state and the number of transitions between states, and expectation is
a linear operator, all we need in the
E-step is to calculate conditional expectations of these two quantities given
the observed data. The conditional expectations are provided in the
following theorem.

\begin{thm}[Conditional expectations \cite{Hobolth2005}]
\label{HJThm}
Consider a continuous time Markov process $\{x(t):0 \leq t \leq T\}$ on
a finite state space with rate matrix $Q$. Denote the transition matrix
$P(t)=\exp(Qt)$. We have the following conditional expectations:
\begin{itemize} 
\item Time spent in state $a$
  \begin{eqnarray*}
    {\rm E}[T(a)|x(0)=i,x(T)=j]=
    \int_0^T P_{ia}(t)P_{aj}(T-t)dt/P_{ij}(T).
  \end{eqnarray*}
\item Number of transitions between states $a$ and $b$
  \begin{eqnarray*}
    {\rm E}[N(a,b)|x(0)=i,x(T)=j]=
    Q(a,b)\int_0^T P_{ia}(t)P_{bj}(T-t)dt/P_{ij}(T).
  \end{eqnarray*}
\end{itemize}
\end{thm}

The transition probability matrix $P(t)=\exp(Qt)$ with entries 
$P_{ab}(T)=P(X(T)=b|X(0)=a)$ is calculated using
an eigenvalue decomposition of~$Q$.
Let $U$ be the orthogonal matrix with eigenvalues as columns and
$D_{\lambda}$ the diagonal matrix of corresponding eigenvectors such that
$Q=UD_{\lambda}U^{-1}$ and
$ P(T)=e^{QT}=Ue^{TD_{\lambda}}U^{-1}.$
In the most general case some of the eigenvalues and eigenvectors are complex.
Conditional expectations are now found from
\begin{eqnarray*}
 \int_0^T P_{ab}(t)P_{cd}(T-t)dt =
 \sum_i U_{ai} U^{-1}_{ib}
 \sum_j U_{cj} U^{-1}_{jd} J_{ij}, \mbox{ where }
 J_{ij}=e^{T\lambda_j} \int_0^T e^{t(\lambda_i -\lambda_j)}dt.
\end{eqnarray*}
Even when the eigenvalues are complex, the $J_{ij}$ integrals are easy
to evaluate.

In the case of multiple sites that are identically distributed and evolve
independently we can summarize the data in a frequency table $\nu(i,j)$
that summarizes the number of sites with $x(0)=i$ and $x(T)=j$. In this case
the complete log-likelihood conditional on the observed data becomes
\begin{eqnarray*}
  G(\theta;\theta^k)=
  \sum_i \sum_j \nu(i,j) 
   {\rm E}_{\theta^k}[\log L(\theta;x)|x(0)=i,x(T)=j].  
\end{eqnarray*}
The linearity in the time spent in a state and the number of jumps between
states is maintained and the EM-algorithm remains simple. Indeed, the 
conditional expectations are given by Theorem~\ref{HJThm} and the M-step for 
the GTR model becomes 
\begin{eqnarray*} 
  \theta_{ab}^*=
  \frac{\sum_i \sum_j \nu(i,j)({\rm E}_{\theta^k}[N(a,b)|x(0)=i,x(T)=j]+
    {\rm E}_{\theta^k}[N(b,a)|x(0)=i,x(T)=j])}
    {\sum_i \sum_j \nu(i,j)
    (\pi_b{\rm E}_{\theta^k}[T(a)|x(0)=i,x(T)=j]+
    \pi_a{\rm E}_{\theta^k}[T(b)|x(0)=i,x(T)=j])}.
\end{eqnarray*}

Note that because we assume $P(t)=\exp(Qt)$, $t$ and $Q$ are confounded.
Indeed, $Qt=(2Q)/(t/2)$, which means that twice the rate at half the time
has the same results. To avoid confounding, the rate matrix is 
calibrated such that time corresponds to expected changes per site.
This means that 
$  \sum_i \sum_{j \neq i} \pi_i Q_{ij}=1.$
\subsection{Multiple sequences}
Now consider the case of multiple sequences related by an unrooted 
phylogenetic tree with $n$ leaves (terminal nodes), $n$ terminal edges, 
$n-2$ internal nodes and $n-3$ internal edges.
The single site complete log-likelihood becomes
\begin{eqnarray*}
  \log L(\theta;x) =\sum_{i=1}^{2n-3}
    \Big( \sum_{a\in \Sigma} T^i(a)Q^i(a,a)+
           \sum_{a\in \Sigma} \sum_{b \neq a} N^i(a,b)\log Q^i(a,b) \Big) 
\end{eqnarray*}
where $T^i(a)$ is the total time spent in state $a$ on edge $i$ and
$N^i(a,b)$ is the number of transitions from $a$ to $b$ on edge $i$.
Letting $y=(y^1,\ldots,y^n)$ be the observed data at the leaves and
letting $a(i),d(i)$ be the ancestral and descendant values at the two ends 
of the edge with descendant node~$i$ we get from the Markov property
\begin{eqnarray*}
  G(\theta;\theta^k) = \sum_{i=1}^{2n-3} \sum_{a(i),d(i)}
  {\rm E}_{\theta^k} \Big[ \sum_{a\in \Sigma} T^i(a)Q^i(a,a)+
    \sum_{a\in \Sigma} \sum_{b \neq a} N^i(a,b)\log Q^i(a,b) |a(i),d(i) \Big]
  P_{\theta^k}(a(i),d(i)|y).
\end{eqnarray*}
Here $P_{\theta^k}(a(i),d(i)|y)$ is the probability of observing the ancestral 
value $a(i)$ and descendant value $d(i)$ at the edge with descendant node~$i$ 
given the data $y$.
These probabilities are calculated using Felsenstein's peeling algorithm 
\cite{Felsenstein1981}.
In the E-step we therefore need to calculate conditional mean values
on each edge. Conditioning on $a(i)$ and $d(i)$, the
conditional mean values are determined by Theorem~\ref{HJThm}

In this paper we consider the case where the rate matrix is the same 
on all edges, but the length varies between edges.
For example, the rate matrix $Q$ could be the GTR rate matrix~(\ref{GTR})
calibrated.
The rate matrix on each edge $i$ is then given by
$  Q^i=w_i Q, \;\; i=1,\ldots,2n-3,$
where $w_i$ is the length of the edge with descendant node~$i$. 
The M-step is slightly more complicated for multiple sequences compared to 
pairwise sequences, but can be carried out as described in 
\cite{Hobolth2005}.
\section{Generalized neighbor-joining}
We will consider binary trees, meaning that all terminal nodes are of 
degree one and all internal nodes are of degree three.
Given a tree, its edge lengths are said to be additive if the distance
between any pair of leaves is the sum of the lengths of the edges on the
path connecting them. Homogeneous finite-state continuous time 
Markov models satisfy
\begin{eqnarray}
  \label{mult}
  \sum_k P_{ik}(t_1)P_{kj}(t_2)=P_{ij}(t_1+t_2).
\end{eqnarray}
Define the maximum likelihood distance \cite{Felsenstein1996} between a
pair of leaves $\{i,j\}$ by 
\begin{eqnarray*}
  d(i,j)={\arg \max}_t \Big\{ \prod_{s=1}^N P_{y^i(s),y^j(s)}(t) \Big\}    
\end{eqnarray*}
where $y^i=(y^i(1),\ldots,y^i(N))$ is the DNA sequence of length $N$ observed
at leaf $i$.
Suppose the leaves $i$ and $j$ are connected by node $k$. 
Using~(\ref{mult}) and consistency of maximum likelihood we get
$d(i,j) \approx t_1+t_2$.
Extending this argument to a general phylogenetic tree we see that maximum
likelihood distances based on continuous time Markov chains between leaf 
sequences should be close to additive if there is enough data to obtain 
reliable estimates of the pairwise distances. 
\subsection{Saitou-Nei neighbor-joining method}
Given a binary tree $\Gamma$ with additive edge lengths, we can
reconstruct it from the pairwise distances $d(i,j)$ of its leaves $\{i, 
\, j\}$ using the neighbor-joining method of Saitou and Nei (1987). 
Firstly, pick a cherry $\{i, \, j\}$ in the tree, i.e. leaves that have the 
same parent node $l$. Secondly, remove the leaves $i$ and $j$ from the set 
of leaves and add $l$, defining its distance to any other leaf $k$ by
\begin{eqnarray*}
  d(l,k)=\frac{1}{2}(d(i,k)+d(j,k)-d(i,j))
\end{eqnarray*}   
and edge lengths $w_i$ and $w_j$ of edges with descendant nodes $i$ and $j$, 
respectively, by
\begin{eqnarray}\label{edgelength}
  w_i &=& \frac{1}{2}\left[ d(i, j) + \frac{1}{n-2}\sum_{k \not = i, k \not = j} (d(i,k)-d(j,k))\right] \\
  w_j &=& d(i,j)-w_i.\label{edgelength2}
\end{eqnarray}
Repeat the procedure until the number of leaves $n$ is 3.
The main problem is picking the cherry, and a solution was suggested in 
Saitou and Nei \cite{Saitou1987} and modified by Studier and Keppler \cite{Studier1988}.  

\begin{thm}[Cherry picking criterion \cite{Saitou1987, Studier1988}]
\label{simplecherry}
Let~$d$ be an additive tree metric and define the $n\times n$-matrix $B$
with entries
\begin{eqnarray*}
  B(i,j)= (n-2) d(i,j) - \sum_{k \neq i} d(i,k) -\sum_{k \neq j} d(j,k).
\end{eqnarray*}
Then the pair of leaves $\{i^*,j^*\}$ that minimizes $B$ is a cherry in the 
tree $\Gamma$.
\end{thm}
\subsection{The generalized neighbor-joining method}
It is natural to ask if we can generalize the neighbor-joining 
method based on pairwise distances to distances 
based on subtrees. Here, a subtree is a set of leaves and a subtree distance 
is the sum of the lengths of the edges on the path connecting the 
subtree leaves.
Distances based on subtrees are expected to be more 
accurately determined than pairwise distances because of the following reasons:
(1) they are calculated from more data, and (2)we include more conditions to 
the calculation. (When we calculate pairwise
distances via MLEs, we assume that each path containing a pair is independent,
while they are not independent because each path share some branches with the 
other.)
Pachter and Speyer \cite{Pachter2004} show that binary trees with additive 
edge lengths are indeed determined from distances based on subtrees.
 Let $\Omega$ be set of leaves in the tree $\Gamma$ and 
let ${\Omega \choose m}$ be the set of all $m$-subsets of $\Omega$.
Then we denote $D(R)$ for $R \in {\Omega \choose m}$ the subtree distance of
the subtree containing $R \in {\Omega \choose m}$.

\begin{thm}[Subtree distance condition \cite{Pachter2004}]
\label{Pachter-Speyer}
Let $\Gamma$ be binary tree with additive edge lengths.
Let $m, \;\; 2 \leq m \leq (n+1)/2,$ be the size of each subtree.
The tree $\Gamma$ is uniquely determined from the set $\{D(R): \, 
R \in {\Omega \choose m}\}$. 
\end{thm}

Pachter and Speyer do not describe how to reconstruct the tree and calculate the 
edge lengths from the subtree distances. 
The tree reconstruction can be found in \cite{Levy2005} and is a two-stage procedure. 

\begin{thm}[Equivalent topologies \cite{Levy2005}]\label{reductionthm}
Suppose that $\{i, \, j\}$ is a pair of leaves in the set of leaves 
$\Omega$ with $n = |\Omega|$ and suppose that $2 \leq m \leq n - 2$. 
  Define the pairwise distances
  \begin{equation}
    \tilde{d}(i,j)= 
    \sum_{\Lambda \in { \Omega \setminus \{i , j\} \choose m-2}}D(\{i,j,\Lambda\}).
  \end{equation}
  The tree $\tilde{\Gamma}$ based on the additive tree metric~$\tilde{d}$ 
  has the same topology as the tree $\Gamma$ based on the additive tree 
  metric~$D$.  
\end{thm}

We can thus find the topology of the tree~$\Gamma$ by using the cherry
picking criterion in 
Theorem~\ref{simplecherry} on the pairwise distances $\tilde{d}(i,j)$.
Using equations (\ref{edgelength}) and (\ref{edgelength2}) we can also find 
the edge lengths $\tilde{w}_i$ for any edge $\tilde{e}_i$ in the tree 
$\tilde{\Gamma}$.
In \cite{Levy2005} the map between the
edge lengths in~$\tilde{\Gamma}$ and the edge lengths in~$\Gamma$ is also 
described.   
The map between the edge lengths in the two metrics is linear and one-to-one 
and given as follows.

\begin{thm}
[Map between edge lengths \cite{Levy2005}]
Let $\Gamma$, $\tilde{\Gamma}$, $m$ and $n$ as in Theorem \ref{reductionthm}.
and let $L_j(e)$ denote the set of leaves in the component of $\Gamma-e$ 
(or equivalently $\tilde{\Gamma}-e$) that
  contains leaf $j$ after removing an edge $e$.
\begin{enumerate}
\item Let $e_i, \; i=n+1,\ldots,2n-3,$ denote the internal edges of $\Gamma$ 
  with lengths $w_i$ and let $\tilde{e}_i$ be the corresponding internal edges 
  of $\tilde{\Gamma}$ with lengths $\tilde{w}_i$. Then
  \begin{eqnarray*}
    w_i = \frac{2\tilde{w}_i} 
    {{|L_a(e_i)|-2 \choose m-2} + {|L_c(e_i)|-2 \choose m-2}},  
  \;\; i=n+1,\ldots,2n-3,  
  \end{eqnarray*}
  where $a$ and $c$ are the leaves on opposite sides of the edge $e_i$.

\item Denote the terminal edges of $\Gamma$ by $e_1,\ldots,e_n$ with 
corresponding edges $\tilde{e}_1,\ldots,\tilde{e}_n$ in $\tilde{\Gamma}$.
Let
\begin{eqnarray*} 
  C_i=\sum_{j=n+1}^{2n-3} \left( {n-2 \choose m-2} - 
  {|L_i(e_j)|-2 \choose m-2} \right) w_j
\end{eqnarray*}  
Then
\begin{eqnarray*}
  \left( \begin{array}{c}
    w_1 \\ \vdots\\ w_n 
  \end{array} \right)  
  = A
  \left( \begin{array}{c}
    2\tilde{w}_1-C_1 \\ \vdots\\ 2\tilde{w}_n-C_n 
  \end{array} \right), \;\;
  {\rm where} \;\;
  A = \frac{1}{2 {n-3 \choose m-2}}
 \left( I - \frac{m-2}{(m-1)(n-2)} J \right).
\end{eqnarray*}
Here, $I$ is the identity matrix and $J$ is the matrix consisting entirely of $1$s.
\end{enumerate}
\end{thm}

The running time for computing the subtree distances is $O(Ln^m)$ where 
$L$ is the length of the alignment and the computation of the distance matrix 
$\tilde{d}$ is $O(n^{m})$ (both steps are trivially parallelizable). 
The subsequent 
neighbor-joining is $O(n^3)$ and edge weight reconstruction is $O(n^2)$. 
It is interesting  to note that for fixed $L$ the running time of the 
algorithm is $O(n^3)$ for both $m=2$ and $m=3$.
\section{The EMGNJ algorithm}
Suppose we wish to estimate the GTR model.  The EMGNJ algorithm consists of
two main parts, reconstructing a tree via the GNJ method and improving
GTR rates via the EM algorithm.  We iterate these steps until it converges.

\begin{algorithm}[The EMGNJ algorithm]
Suppose we have $n$ DNA sequences and an integer $2 \leq m \leq n - 2$.
\vskip 0.2in

\noindent
{\bf Input}: $n$ DNA sequences and an integer $2 \leq m \leq n - 2$.

\noindent
{\bf Output}: The GTR rates and a phylogenetic tree.

\begin{enumerate}
  \item\label{S1} Estimate stationary distribution from empirical frequencies.
  \item\label{S2} Reconstruct tree using MJOIN under the Jukes-Cantor (JC69) 
   model.
  \item\label{S3} Estimate GTR substitution rates and edge lengths from 
   current tree.
  \item\label{S4} Reconstruct tree using MJOIN and current GTR rates.
  \item\label{S5} If likelihood is not improved return current tree and GTR 
    rates;
    otherwise go to 3.
\end{enumerate}
\end{algorithm}
In other words, we provide starting values of the algorithm in Step \ref{S1}
and Step \ref{S2} and iterate Step \ref{S3} and  Step \ref{S4} until 
convergence.
\section{Computation}
We implemented subroutines of the EMGNJ algorithm, Step \ref{S3} and 
Step \ref{S4} with $m = 4$ under the JC model.  
We plan that a full implementation of the algorithm will be released soon. 
We applied our implementation to find the phylogenetic tree for 21 {\em S-locus} 
receptor kinase (SRK) sequences from \cite{Sainudiin2005a} involved in the self/nonself discriminating self-incompatibility 
system of the mustard family described in \cite{Nasrallah2002}.

\begin{table}
\caption{\label{symmT}Symmetric difference ($\mathbf{\Delta}$) between $10,000$ trees sampled from the 
likelihood function via MCMC and the trees reconstructed by 5 methods. sub-EMGNJ means the implementation of subroutines, Step \ref{S3} and Step \ref{S4}.}
\centering
\begin{tabular}{|r|r|r|r|r|r|r|}\hline
$\mathbf{\Delta}$ & sub-EMGNJ & Saitou-Nei NJ & fastDNAml 	& DNAml(A) 	& DNAml(B) 	& TrExML\\ \hline 
0  & 	0  &		0 	&  	0	&	2	&	3608	& 	0\\
2  & 		77 &	0 	&  	0 	&	1	&	471	&	0\\
4  & 	3616 &		171	& 	6	&	3619	&	5614	&	0\\
6  & 	680 &		5687	&	5	&	463	&	294	&	5\\
8  & 		5615 &		4134	&	3987	&	5636	&	13	&	71\\
10 & 		12 &		8	&	5720	&	269	&	0	&	3634\\
12 & 		0 &		0	&	272	&	10	&	0	&	652\\
14 & 		0 &		0	&	10	&	0	&	0	&	5631\\
16 & 		0 &		0 	&	 0 	&	0	&	0	&	7\\ \hline
\end{tabular}
\end{table}
We sampled $10,000$ trees from a Markov chain with stationary distribution proportional 
to the likelihood function by means of a Markov chain Monte Carlo (MCMC) algorithm implemented in 
{\tt PHYBAYES} \cite{Aris-Brosou2003}.  
We then compared the tree topology of 
each tree generated by this MCMC method with that of the 
reconstructed trees via Step \ref{S3} and Step \ref{S4} in the EMGNJ method, Saitou-Nei NJ method, {\tt fastDNAml}, {\tt DNAml} 
from {\tt PHYLIP} package by \cite{Felsenstein2004}, and {\tt TrExML} 
\cite{Wolf2000} under their respective default settings with the JC69 model.  
We used {\tt treedist} from {\tt PHYLIP} to compare two tree topologies.  If the 
symmetric difference $\Delta$ between two topologies is $0$, then the two topologies are 
identical.  Larger $\Delta$'s are reflective of a larger distance between the two compared 
topologies.  Table \ref{symmT} summarizes the distance between a reconstructed tree and the 
MCMC samples from the normalized likelihood function.  For example, the first 
two elements in the third row of Table \ref{symmT} mean that $171$ out of the $10,000$ MCMC sampled trees 
are at a symmetric difference of $4$ ($\Delta = 4$) from the tree reconstructed via Saitou-Nei NJ method (with pairwise distance).  
DNAml was used in two ways: 
DNAml(A) is a basic search with no global rearrangements, whereas DNAml(B) applies a broader search with 
global rearrangements and $100$ jumbled inputs.  The fruits of the broader search are reflected by the 
accumulation of MCMC sampled trees over small $\Delta$ values from the DNAml(B) tree.  Note that the tree topology for the reconstructed tree via Step \ref{S3} and Step \ref{S4} in the EMGNJ method and one via the GNJ method ({\tt MJOIN}
\cite{Levy2005}) have the same tree topology.

\section{Conclusion}
We expect to improve the results using a full implementation of the EMGNJ method
for two main reasons:
(1) Even though currently we have an implementation with subroutines 
Step \ref{S3} and Step \ref{S4}, the result is improved compared to Saitou-Nei
NJ method and to {\tt fastDNAml} as in Table \ref{symmT}  
(2) Iterating Step \ref{S3} and Step \ref{S4} should improve the GTR rates.

We want to investigate the number of iterations until the output from
the EMGNJ method converges.  
It is also of interest to study the results with different values of $m$. 
\section{Discussion}
A potential problem is that the EM algorithm often get stuck in local maxima.
We might be able to avoid this problem using Moore Rejection Sampling 
\cite{Sainudiin2005c}.
We sample data via the sampler for each subtree and we reconstruct
trees with a set of samples under the JC69 model.  
Then we estimate the GTR rates and trees 
via the EMGNJ method.  At the end, we compare the likelihood values of these
trees and we take the tree with the biggest likelihood value.
One notices that the process to reconstruct trees via the EMGNJ method 
from samples can be parallelizable.
\section*{Acknowledgement}
AH acknowledges financial support from the Danish Research Council 
(Grant 21-04-0375). 


%

\end{document}